# Fast Scanning Probe Microscopy via Machine Learning: Non-rectangular scans with compressed sensing and Gaussian process optimization

Kyle P. Kelley,[1] Maxim Ziatdinov,[1] Liam Collins,[1] Michael A. Susner,[2] Rama K. Vasudevan,[1] Nina Balke,[1] Sergei V. Kalinin,[1] Stephen Jesse[1]

[1]The Center for Nanophase Materials Sciences, Oak Ridge National Laboratory, Oak Ridge, Tennessee 37831, United States

[2]Materials and Manufacturing Directorate, Air Force Research Laboratory, Wright-Patterson Air Force Base, Ohio 45433, United States

Abstract:

Fast scanning probe microscopy enabled via machine learning allows for a broad range of nanoscale, temporally resolved physics to be uncovered. However, such examples for functional imaging are few in number. Here, using piezoresponse force microscopy (PFM) as a model application, we demonstrate a factor of 5.8 improvement in imaging rate using a combination of sparse spiral scanning with compressive sensing and Gaussian processing reconstruction. It is found that even extremely sparse scans offer strong reconstructions with less than 6 % error for Gaussian processing reconstructions. Further, we analyze the error associated with each reconstructive technique per reconstruction iteration finding the error is similar past approximately 15 iterations, while at initial iterations Gaussian processing outperforms compressive sensing. This study highlights the capabilities of reconstruction techniques when applied to sparse data, particularly sparse spiral PFM scans, with broad applications in scanning probe and electron microscopies.

The advent of scanning probe microscopy (SPM) techniques have led to extraordinary advances in our understanding of materials and devices on the nanoscale, enabling a nanoscience revolution over the last three decades.[1,2] SPM tools have had a major impact on multiple areas of science and technology, ranging from structural mapping across mesoscopic and atomic scales[3,4] visualizing solid-liquid interfaces,[5,6] quantifying nanomechanical properties,[7-11] as well as probing electrical,[12-14] ferroelectric,[15-19] and magnetic[20-22] functionalities but to name a few. In certain cases, the application of SPM techniques helped to usher in new areas of science. Notable examples include 1) single molecule unfolding spectroscopy[23] which opened the window for exploring thermodynamics and kinetics of single-molecule biochemical reactions using desktop instrumentation, and 2) piezoresponse force microscopy and spectroscopy[24-27] that enabled quantitative studies of polarization dynamics and domain structures in ferroelectric[28], multiferroic,[29] and ionic materials[30,31] for applications ranging from information technology devices to batteries[32] and fuel cells[33].

Yet despite the proliferation of SPMs in virtually all areas of modern experimental sciences, the basic paradigm of SPM remains invariant- the raster scanning of surfaces with detection of response signal (or spectral data sets) on a rectangular grid. It has long been argued that rectangular scans with uniform spatial sampling are best suited for the exploration of *a priori* unknown surfaces, whereas the presence of spatially localized objects of interest prompt the development of spatially non-uniform grids in automated experiments. Similarly, rectangular scanning grids at constant speed result in the high-nonuniformity of the accelerations experienced by the moving probe and thus limit the maximal scanning speeds.

As such, the broad adoption of non-rectangular scans[34-40] necessitates the development of the algorithmic tools that allow conversion of the data stream acquired along the fixed or dynamically adjusted probe path on the classical rectangular grids. It should be noted that most of the non-rectangular scans were motivated by the need to accelerate the scanning speed and were generally either spiral or Lissajous scans. At the same time, more complex scan paths to follow specific geometric objects have also been reported.[41] Similarly, incorporating sparsity in the scanning geometry enables a number of benefits (e.g. minimized perturbation to the system, extended probe life, less acquisition pixels) arising from the reduced amount of tip-sample interactions. These benefits are particularly useful for techniques, such as piezoresponse force microscopy (PFM), nanomechanical mapping, and Kelvin probe force microscopy, where changes to the probe or sample due to the intrinsic methodology are detrimental to the experiment.[42-51] With further optimization and understanding of errors associated with sparse scanning geometries, these techniques would be of extreme interest to a broad range of practitioners in nanoscience and nanotechnology.

Accordingly, we present two methods for reconstructing sparse PFM spiral scans: (i) compressive sensing[52-59] (CS) and (ii) Gaussian processes[60-64] (GP) regression. The compressive sensing image inpainting algorithm requires two inputs: (i) The sparse and noisy measurements $\mathbf{y}$, and (ii) the scanning mask indicating sampled pixel locations. To recover the full image $\boldsymbol{f_0}$ from $\mathbf{y}$

$$\mathbf{y} = \Phi \boldsymbol{f_0} \quad (1)$$

We particularly rely on redundant wavelet frames to recover an image from the sparse measurements. Image reconstruction is achieved by estimating the coefficient $x$ in the wavelet basis

$$\operatorname{argmin}_\mathbf{x} = \frac{1}{2}\|\mathbf{y} - \Phi \mathbf{W}\mathbf{x}\|^2 + \lambda\|\mathbf{x}\|_1 \quad (2)$$

where $\boldsymbol{f}$ is the reconstructed image, $\mathbf{W}$ is the wavelet synthesis operator, and $\lambda$ is the tuning parameter of sparsity. The soft thresholding operator, $S_\lambda^W$, in wavelet basis is defined as

$$S_\lambda^W(\boldsymbol{f}) = \sum_m S_\lambda(\langle \boldsymbol{f}, \mathbf{w_m}\rangle)\mathbf{w_m} \quad (3)$$

where $\mathbf{w_m}$ is the basis of the wavelet frame, and

$$S_\lambda(x) = x * \max(0, 1 - \frac{\lambda}{|x|}) \quad (4)$$

During each iteration, the reconstruction is updated as

$$\boldsymbol{f}^{(i+1)} = S_\lambda^W\left(\operatorname{ProjC}(\boldsymbol{f}^{(i)})\right) \quad (5)$$

where $\operatorname{ProjC}(\boldsymbol{f})$ is the gradient descent of the data fidelity term:

$$\operatorname{ProjC}(\boldsymbol{f}) = \mathbf{M}\boldsymbol{y} + (1 - \mathbf{M})\boldsymbol{f} \quad (6)$$

In this work, we perform a serial implementation of the code by Li et al. (Li, 2018, MM).[1*] Generally, compressive sensing is an elegant method for obtaining solutions to an unknown system and allows for data reconstruction from fewer points than is required by the Nyquist-Shannon sampling theorem.[65] However, limitations such as reconstruction accuracy necessitate the need for additional reconstruction methods.

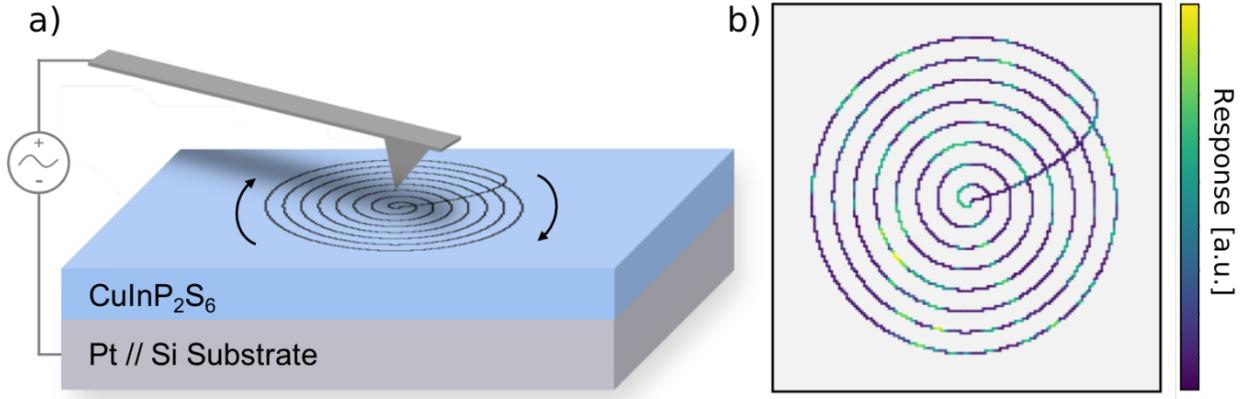

**Figure 1: Illustration of spiral scan technique.** (a) PFM spiral scanning illustration of $CuInP_2S_6$ van der Waal crystal mechanically exfoliated on a Pt//Si substrate. (b) Corresponding raw data collection from spiral based scanning illustrated in Figure 1a.

As an alternative approach, in GP regression, one estimates an unknown non-linear function from noisy observations of this function at a finite number of points, assuming that the observations are a sample from the multivariate Gaussian distribution. The observations at different locations are assumed to be linked via the covariance function (kernel) whose parameters can be learned from the data as a part of the regression process. GP regression allows using information from a finite number of sparse measurements and the learned kernel parameters to predict the function value in the unexplored locations on a sample and provides an uncertainty for those predictions. Here we utilized a sparse GP approach, which uses a global GP interpolation on an underlying base kernel to create an approximate kernel for tractable computations. Specifically, because all our measurements lie on a regularly spaced grid, we were able to wrap a radial basis function (RBF) kernel into a grid kernel ,[66] which exploits Toeplitz and Kronecker structure within the covariance matrix for dramatically faster computations.

Here, we experimentally use the aforementioned reconstruction techniques on sparse PFM scans with Archimedean based spirals employed in a standard AFM configuration, as seen in Figure 1. Using a Cypher AFM (Asylum Instruments an Oxford Instruments Company) in conjunction with a LabView based framework, spiral scans spanning approximately 3-5 ums in diameter were measured in PFM. Furthermore, images were acquired at a frame rate of 0.25 frames/s with a constant angular velocity. Spiral images were taken using 62, 32, 24, 16, and 8 spirals with the basic implementation shown in Figure 1b. Correspondingly, compressed sensing and Gaussian processing-based models were used for image reconstruction and compared in terms of reconstruction veracity. This approach allows for varying degrees of sparsity to be utilized in fast scanning SPM, providing a foundation for ultra-fast SPM scanning.

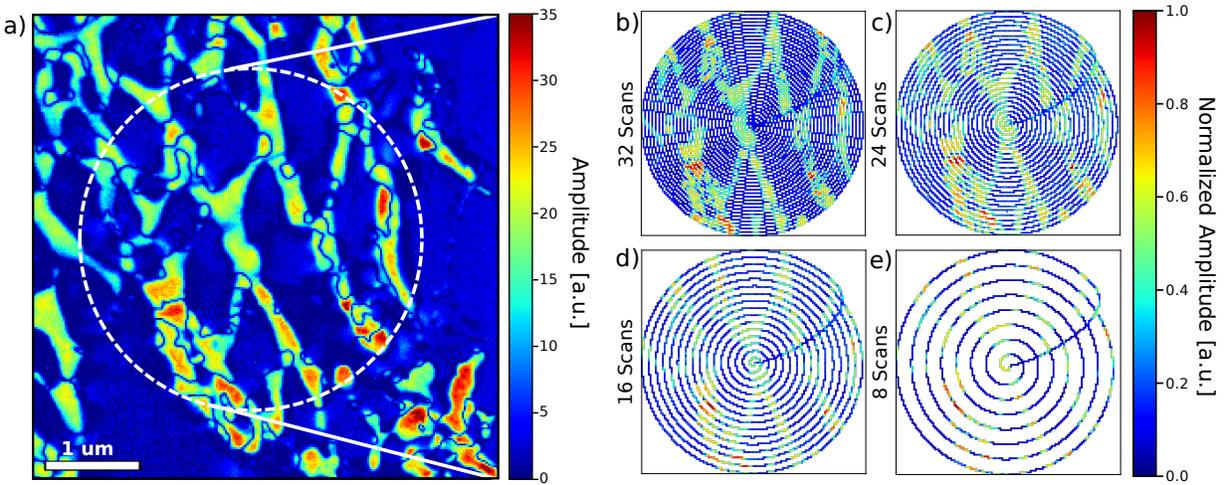

**Figure 2: PFM with varying data sparsity. (a)** Representative band excitation PFM amplitude of $CuInP_2S_6$ illustrating platform for compressive sensing and Gaussian processing based reconstructive techniques. **(b-e)** Normalized amplitude data from Archimedean based spiral scans spanning 32, 24, 16, and 8 spirals, respectively.

As a model material system, we choose single crystals of $CuInP_2S_6/In_{4/3}P_2S_6$ self-assembled heterostructures (CIPS/IPS) as it exhibits localized phase-separated ferroelectric domains, providing an ideal platform for employing reconstructive based techniques. Specifically, CIPS is a Van der Waals crystal where mobile Cu ions are segregated into regions corresponding to ferroelectrically active areas on the order of 500 nm. Typical samples are comprised of a ferroelectric CIPS matrix with embedded paraelectric IPS islands. Within these ferroelectric regions, four different polarization states with distinctively different piezoelectric properties are accessible producing not only domain walls, but also contrast within these domains, as seen in Figure 2a.[67] The zero-amplitude areas represent a non-ferroelectric $In_{4/3}P_2S_6$ phase which we use as an internal reference point. This phase is formed after phase separation in non-stochiometric $Cu_{0.4}In_{1.2}P_2S_6$.[68] Further details concerning the ferroelectric properties of CIPS can be found elsewhere.[69-73] Nevertheless, CIPS with feature sizes spanning orders of magnitude is an excellent basis for testing sparse data reconstruction.

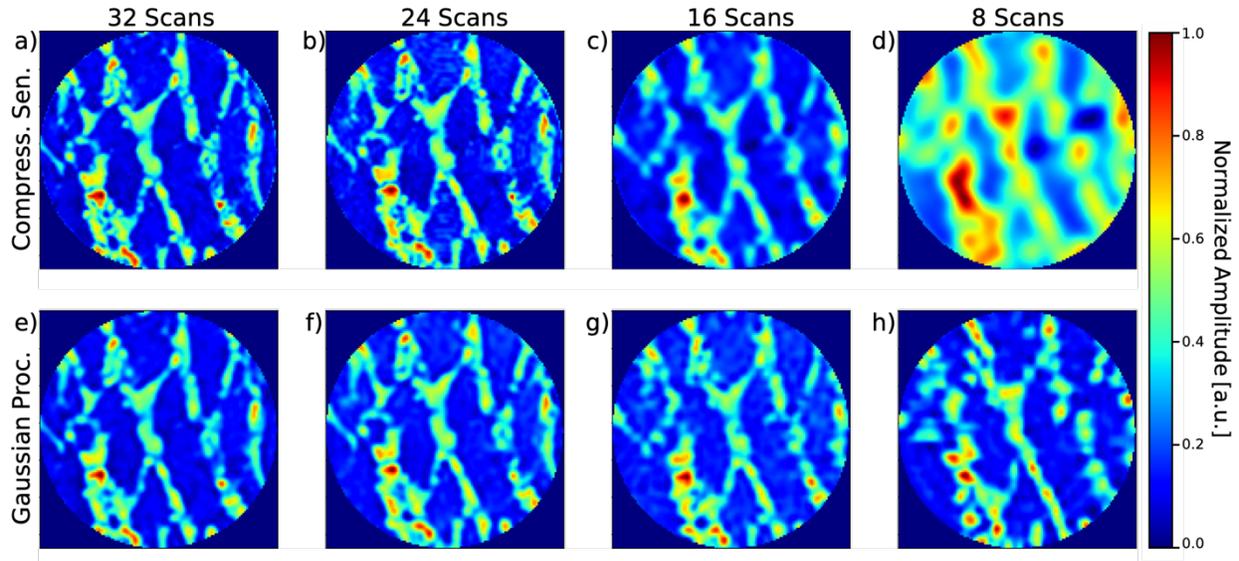

**Figure 3: Sparse image reconstructions. (a-d)** PFM amplitude reconstruction using compressive sensing and **(e-h)** Gaussian processing algorithms for 32, 24, 16, and 8 spirals (column 1-4 respectively). Images are represented as normalized amplitude calculated from respective reconstruction techniques.

Shown in Figure 2 is a representative PFM amplitude response collected using band excitation[74] PFM with regular raster scanning (Figure 2a) and single frequency PFM spiral scans with different levels of sparsity. The white dotted inset in Figure 2a indicates the region where spiral scans were engaged (Fig 2(b-e)). Note that the spiral scans are presented as normalized PFM amplitude to allow for direct comparison between different density of spirals. At 32 scans (Fig. 2b), the spiraled single frequency PFM scan clearly shows a resemblance to the dotted area seen in Figure 2a where detailed features are still present, namely the high response regions and domain walls. However, as the scan number is reduced to 8 scans (Fig. 2e), the detailed features are virtually unobservable, and the image is less interpretable. Here, we employ reconstructive models to recover the image with finite degrees of accuracy and error.

The results for CS (Fig. 3(a-d)) and GP (Fig. 3(e-h)) regression reconstruction of several sparse measurements are shown in Figure 3. For measurements with small sparsity (small number of data points missing), particularly scans with 32 and 24 spirals, the results for compressive sensing and GP regression are similar. However, for the images with larger sparsity (i.e. a larger number of missing points such as the scans taken with 16 and 8 spirals), the GP tends to outperform CS in making "sharper" predictions.

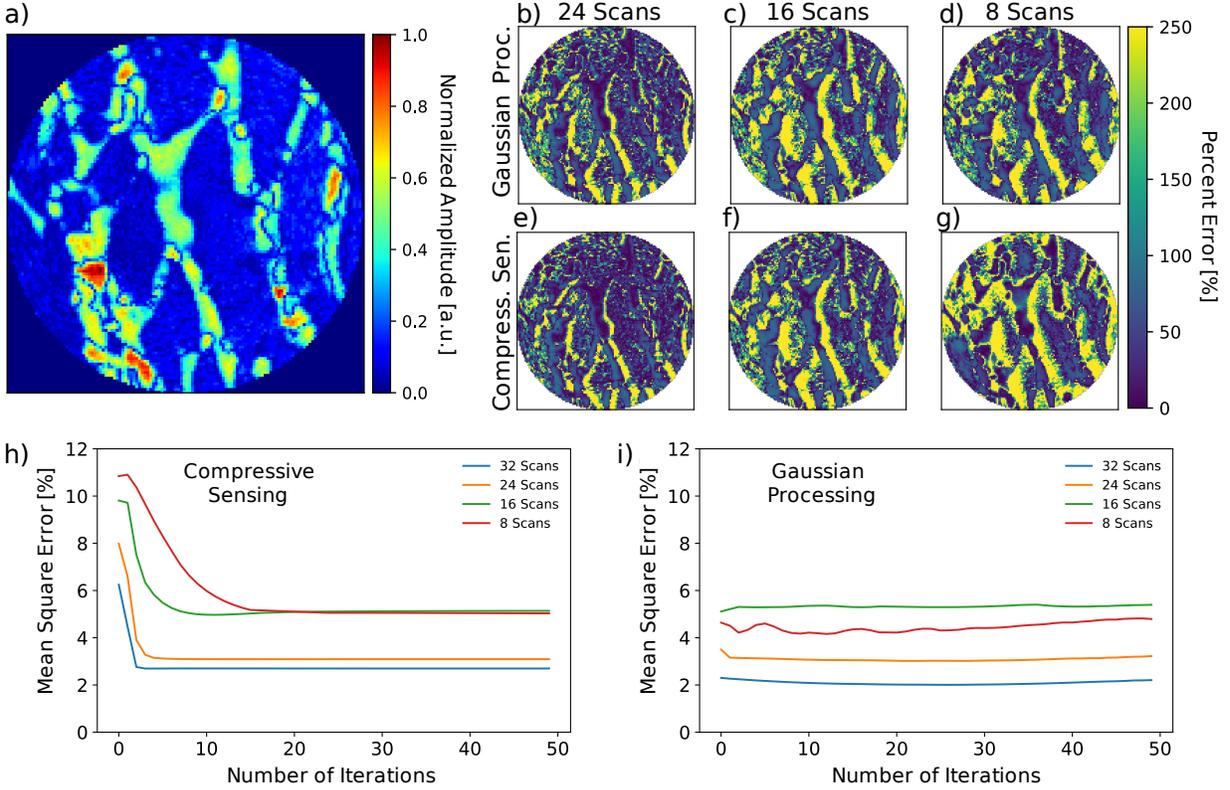

**Figure 4: Error analysis for reconstruction techniques vs number of scans and iterations. a)** Ground truth image constructed using 64 scans with normalized PFM amplitude scale, **(b-d)** Gaussian processing and **(e-f)** compressive sensing reconstruction error at 50 iterations for 24, 16, and 8 scans, respectively. Mean square error as function of reconstruction iterations for compressive sensing **(h)** and Gaussian processing **(i)**.

To understand the errors associated with each reconstruction technique, Figure 4 shows a comprehensive error analysis with percent error at 50 iterations (Fig. 4(b-g)) and mean square error versus number of iterations for scans with 24, 16, and 8 spirals (Fig. 4(h,i)). The ground truth basis for calculating the associated error for each reconstruction technique is a scan with 64 spirals where 'all' data points are captured (Fig. 4a). CS shows a qualitative increase in percent error as the number of spirals is decreased from 24 to 8. In contrast, GP regression percent error seems to be similar at 16 and 8 spirals. Interestingly, all spiral scans for both CS and GP regression reconstructions have an asymmetric percent error located on the right side of the ferroelectric domains, or high intensity areas. We further calculate global error associated with each reconstruction technique to understand the dependence on reconstruction iterations, Figure 4(h,i).

Generally, GP regression outperforms CS in both the number of spirals and the number of iterations. However, there are 2 distinct features represented in the mean square error: (i) CS reconstruction of 16 and 8 scans asymptotically approach a mean square error of ~ 5%, (ii) GP reconstruction of 8 scans (4.4% error on average) has a smaller mean square error value compared to that of 16 scans (5.1% error on average). It is noteworthy that reducing the number of spirals from 64 (12869 pixels) to 8 (2194 pixels) results in a factor of 5.8 reduction in collected data, enabling an increase in scan speed, with less than 5% error. Ultimately, these results confirm fast scanning probe microscopy enabled via machine learning as a viable pathway for exploring temporally resolved phenomenon.

In summary, here we demonstrate a method to increase acquisition speed of SPM utilizing a combination of spiral scanning in conjunction with image reconstruction techniques. Using sparse spirals ranging from 64 to 8 spirals over the same spatial location, we observe the reconstruction error of two methods, compressed sensing and Gaussian process regression, and find that Gaussian regression more accurately capturing the ground truth, although it is computationally more expensive. Further developments of these methods are possible through implementation of additional machine learning, such as neural networks to better inform acquisition and the Gaussian processes inducing points. This approach is applicable to a diverse set of disciplines in nanoscience and nanotechnology where attempts to increase the temporal resolution, or to lower overall excitation dosage and tip-sample interactions, are needed.

**ACKNOWLEGEMENTS:**

The work was supported by the U.S. Department of Energy, Office of Science, Materials Sciences and Engineering Division (K.P.K., R.K.V., N.B.). The PFM and image reconstruction work was conducted at and supported by the Center for Nanophase Materials Sciences, which is a DOE Office of Science User Facility (S.V.K, L.C. ). Partial support for sample synthesis was provided by the Laboratory Directed Research and Development program at the Oak Ridge National Laboratory (Michael A. McGuire). A portion of the writing of this manuscript was supported by AFOSR (LRIR # 16RXCOR322) and AFRL/RX (Lab Director's Funds).

## CONFLICTS

The Authors declare no Competing Financial or Non-Financial Interests.

## CONTRIBUTIONS

K.P.K and L.C. performed the experimental measurements; M.S. grew the bulk crystals; M.Z. and R.K.V developed image reconstruction code; S.J. developed data acquisition system; All authors participated in discussions and contributed in finalizing the paper.

## MATERIALS AND METHODS:

### Materials:

$CuInP_2S_6$ flakes were mechanically exfoliated onto a platinum coated silicon substrate with a thickness greater than 100 nm. The average composition of the flakes was $Cu_{0.4}In_{1.2}P_2S_6$. Additional material details can be found elsewhere. [75]

### Instrumentation:

Spiral scanning geometries were employed via a customized Oxford Instruments Cypher atomic force microscope. Real time signal processing and motor controls were driven by a National Instruments USB-7856R multifunctional RIO FPGA and a combined MATLAB-python platform. Piezoresponse force microscopy (PFM) measurements with spiral geometries were implemented through the Igor-Cypher software with an AC scanning voltage of 1V. Similarly, band excitation PFM measurements were collected using an AC scanning voltage of 1V. Details describing band excitation functionality can be found elsewhere.[74] All experiments used Budget Sensor Multi75E-G Cr/Pt coated AFM probes (~ 3N/m).

### Sparse image reconstruction techniques:

Details outlining the compressive sensing and Gaussian process reconstructive techniques can be found at https://git.io/JfeQr

## DATA AVAILABLITY

The datasets generated during and/or analyzed during the current study are available from the corresponding author on reasonable request

# References

[1*]    We note that the reconstruction quality can be further improved by the GPU implementation in (Li, 2008 MM), where the computation takes milliseconds (< 10 ms) for an image of size 256 by 256.